\documentstyle[draft]{mn}
\begin{document}
\input{epsf.sty}
\title[The impact of HH2 on local dust and gas]
{The impact of HH2 on local dust and gas}

\author[W. R. F. Dent, R. S. Furuya, C. J. Davis]
       {W.R.F.Dent$^1$, R.S.Furuya$^2$, C.J.Davis$^3$, \\
 $^1$UK Astronomy Technology Centre, Royal Observatory, 
Blackford Hill, Edinburgh EH9 3HJ, Scotland\\
 $^2$ INAF, Observatorio Astrofisico di Arcetri,
Largo Enrico Fermi 5, I-50125, Firenze, Italy \\
 $^3$ Joint Astronomy Centre, 660 N. Aohoku Place, Hilo,
Hawaii 96720, USA
}


\pagerange{000--000}
%
%
%
%
\def\cm{\,{\rm cm}}
\def\cc{\,{\rm cm^{-3}}}
\def\asec{\,{\rm ''}}
\def\amin{\,{\rm '}}
\def\kps{\,{\rm {km~s^{-1}}}}
\def\Msun{\,{\rm M_{\odot}}}
\def\Lsun{\,{\rm L_{\odot}}}
\newcommand{\degree}{\mbox{\,$^\circ$}}        
\newcommand{\micron}{\mbox{\,${\mu}$m}}        
\newcommand{\elec}{\mbox{\,e$^{-}$}}           
\def\elec{e$^{-}$}
%
\def\aa{{A\&A}}
\def\aar{{A\&ARev}}
\def\aas{{A\&ASup}}
\def\aj{{AJ}}
\def\apj{{ApJ}}
\def\apjl{{ApJLet}}
\def\apjs{{ApJSup}}
\def\araa{{ARAA}}
\def\ass{{A\&Sp.Sci.}}
\def\mnras{{MNRAS}}
\def\pasp{{PASP}}
\setlength{\topmargin}{-10mm}
\maketitle
\begin{abstract}

We present results from a study of molecular gas and dust in the
vicinity of the Herbig Haro object HH2. Emission from the sub-mm continuum,
$^{12}$CO and HCO$^+$ was mapped
with angular resolutions ranging from 14$\asec$
to 5$\asec$ (or 0.01~pc at the distance of HH2).
The continuum shows an extended dust clump of mass 3.8M$_\odot$
and temperature 22K, located
downstream of the bright optical HH knots. However, a compact
emission peak lies within 0.01~pc
of the low-excitation H$_2$-prominent shocks, with
a luminosity consistent with local heating by the outflow.

The HCO$^+$ emission shows two velocity components: firstly,
ambient-velocity gas lying in a region roughly
corresponding to the dust clump, with abundance enhanced by a factor
of a few close to the H$_2$-prominent knots.
Secondly a component of high-velocity emission (20 $km s^{-1}$ linewidth),
found mainly in a collimated jet linking the low-excitation HH objects.
In this high-velocity jet, the line wings show an abundance ratio
$\chi_{HCO^+}/ \chi_{CO} \propto v^2$, with an HCO$^+$
enhancement compared with
ambient gas of up to $\sim 10^3$ at the most extreme velocities.
Such high abundances are consistent with models of shock chemistry
in turbulent mixing layers at the interaction boundaries of jets.
Extrapolating this effect to low velocities suggests that the more modest
HCO$^+$ enhancement in the clump gas
could be caused by low velocity shocks. A UV precursor may not, therefore
be necessary to explain the elevated HCO$^+$ abundance in this gas.

\end{abstract}

\vskip 5mm

\vskip 10mm

\section{Introduction}

The Herbig-Haro objects HH1 and HH2 are two of
the brightest optical shocks associated with outflows from young stars.
Separated by 140$\asec$ or 0.3~pc at the distance of 460~pc,
both their proper motions and the detection of a faint jet indicate
they are driven by
the embedded 32$\Lsun$ Class I young star known as VLA1
(Herbig \& Jones, 1981; Pravdo et al., 1985).
Because of their brightness, HH1 and 2 have been the subject of intense
study, particularly in optical and near-infrared lines used to
tracing shocked gas (eg Hester et al., 1998). The images show complex
and sometimes bewildering structures which, in the case of HH1, are often
interpreted as multiple overlapping bow shocks.
HH2, however, shows an apparently more random set of knots,
although it too has a number of mini-bow shocks identified through
infrared images (eg Davis et al., 1994).
In order to aid orientation in the region, in Figure 1 we show a near-IR
H$_2$ image of the cluster of
shock fronts that comprise HH2 (from Davis et al. 1994). Individual
features and knots are labelled.

\epsfverbosetrue
\epsfxsize=8.0 cm
\begin{figure}
\center{
\leavevmode
\epsfbox{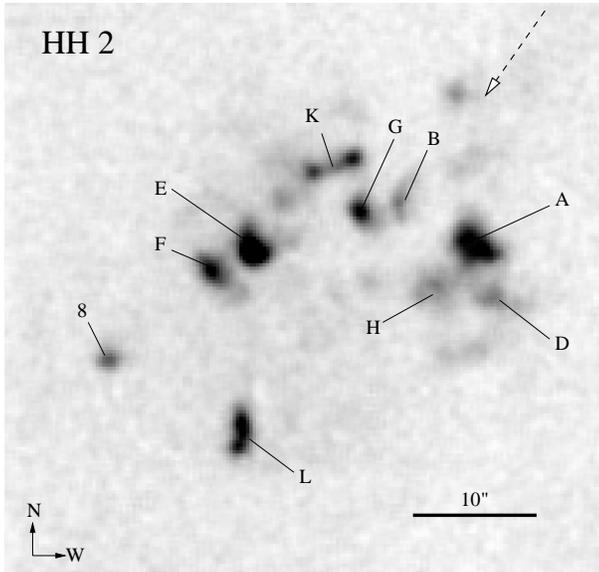}
\caption { 
Image of H$_2$ emission from the vicinity of HH2, based on data
from Davis et al. (1994). The individually-identified knots are labelled and
the flow direction is indicated by the dashed arrow in the upper right. Note
that ``Knot 8'' is only seen in H$_2$ emission (it has no known optical
counterpart).
}
}
\end{figure}

In addition to HH1 and 2, a pair of much
larger, more diffuse HH objects HH401/402 can be seen
almost $20 \arcmin$ (2.6~pc)
further from VLA1 at the same position angle (Ogura, 1995). Thus
HH1 and 2 may actually be inner knots in a ``parcsec-scale'' outflow.

Single-dish HCO$^+$ observations in the immediate vicinity of HH2 have
shown an emission peak downwind of the optical knots, and it was suggested
that this could be due to enhancement in the HCO$^+$
abundance in the ambient gas
(Davis et al., 1990; Dent 1997). Similar HCO$^+$ peaks have been found 
in several other outflows (eg Rudolph \& Welch, 1992). A common
feature is that the emission lies close to ambient velocity, and
the peak appears ahead of the HH shocks. It was proposed
that these clumps are
caused by UV photons from the HH shock inducing
non-equilibrium chemistry in ambient material
in either purely the gas phase (eg Wolfire \& K\"onigl, 1993),
or after the release of icey grain mantles (eg Taylor \& Williams, 1996;
Viti \& Williams, 1999). It is also possible for HCO$^+$ in
entrained shocked gas to be chemically enhanced by the shock
itself (eg Taylor \& Raga, 1995).
Further imaging of HH1 and 2 showed compact clumps of NH$_3$
in the vicinity (Torrelles et al., 1992), again
thought to be due to changes in the molecular abundance. A more extensive
survey of other species towards the HCO$^+$ peak
has shown abundances consistent with the effect
of UV photons from the nearby HH shock (Girart et al., 2002).

The molecular outflow associated with HH1-2 is not prominent, but line wings
have been detected in CO towards HH2, albeit at relatively low velocities
(eg Correia et al., 1997; Dent, 1997).
Maps of the line wing emission over the whole region show a bipolar
outflow at certain velocity ranges,
which appears to terminate near HH1 and HH2; it was suggested that the
flow axis is inclined only 5-10\degree from the plane
of the sky (Moro-Mart\'in et al., 1999).

In order to investigate the relationship between the ambient clump and
high-velocity gas, we have mapped the region in the sub-mm continuum
and J=3-2 CO line, and also obtained higher-resolution
J=1-0 HCO$^+$ images.

\section{Observations}

The continuum observations were made using SCUBA on the JCMT in June 1998.
Imaging was carried out at 850$\micron$
using a standard 64-position fully-sampled jiggle map,
and data were calibrated using Mars, resulting in a calibration
accuracy of $\sim 10\%$.
A fully-sampled map of the J=3-2 $^{12}$CO line was also obtained
using the JCMT; additional J=3-2 spectra of the
isotopomers $^{13}$CO and C$^{18}$O
were extracted from the JCMT Archive. These data were taken in 1994,
using the receiver RxB3i and the DAS spectrometer with a spectral resolution
of $0.3 km s^{-1}$. The beamsize of the JCMT in both these continuum and
spectral line observations was 14$\asec$, and estimated pointing
uncertainties were less than 2$\asec$.

We conducted interferometric imaging of HCO$^+$ J=1-0 line emission
(rest frequency 89.18852 GHz) using the
6-element Nobeyama Millimeter Array (NMA) in 1999
March and May. The primary beam size (field of view) is 83$\asec$
(fwhm) at 89 GHz.
Observations were performed with the C and D configurations, and the resulting
synthesized beam size was $6.9\arcsec\times 4.6\arcsec$ at
P.A.$=-20^{\circ}$. Since the minimum projected baseline length was
3.9 k$\lambda$,  our observations were insensitive to structure extended
by more than $\sim 52\asec$. The phase tracking center was set on the position of the
peak of J=3-2 HCO$^+$
($5^h~36^m~26.91^s, -6^{\circ}~47\amin ~31.9\asec$; 2000.0).
We employed SIS receivers which had double sideband
system noise temperatures of 200-300
K toward zenith.  We used the high spectral resolution FX correlator
for the HCO$^+$ emission and the Ultra-Wideband
Correlator for calibration using continuum emission.
The FX correlator provides a velocity resolution of 0.1 km s$^{-1}$ and total
velocity coverage of 108 km s$^{-1}$ at this frequency. We used 3C273
as the bandpass calibrator and 0528+134 as a phase and gain calibrator.
By comparing with Uranus through 3C84 and 3C454.3,
we estimated the flux density
of 0528+134 to be 2.5Jy during both observing periods,
with an estimated 10\% uncertainty. Imaging was
performed using the AIPS package to make clean maps. The typical 
resulting rms noise
levels were 75 mJy beam$^{-1}$ with a resolution of $0.1 km s^{-1}$.

\section{Results}

\subsection{Dust continuum}

The emission from HH2 observed at 850$\micron$ is shown by the contour map in
Figure~2.
For comparison purposes, the shocked H$_2$ image from Davis et al. (1994)
is superimposed as a greyscale. These H$_2$ knots
lie towards the end of the Southeast lobe
of the outflow; the flow origin (VLA~1) is
situated some $\sim 90 \asec$ to the Northwest.
An extended region of sub-mm continuum
emission, of dimensions $60\asec \times 40 \asec$ (0.13$\times$0.09~pc, fwhm)
is clearly seen downstream from most of the
HH objects. The peak in sub-mm emission lies at $5^h36^m27.0^s,
-6^{\circ}47\amin 21\asec$ (2000.0), and
could be either due to a local maximum in the dust
temperature or density. The J=4-3 HCO$^+$ line observed with the same spatial
resolution peaks $10\asec$ to the south (Dent, 1997), and so the positions are
consistent within the estimated errors.
Girart et al. (2002) also found that
the SO $3_2 - 2_1$ line reaches a peak at this position.
Although there are at least 3 bright sub-mm continuum
clumps within 2 arcmin of
VLA~1 (Chini et al., 2001), most of these have evidence of
an embedded young star.
The dust clump near HH2 has no such known object, even though
the extinction through the cloud is relatively low (see Girart et al.,
2002). The close association with the optical, infrared
and HCO$^+$ emission suggests that the sub-mm continuum
is tracing a cloud closely related to the HH objects.
Furthermore Fig.~2 shows that at least two faint HH knots are apparently
embedded within the sub-mm cloud; the significance of this will be
discussed later. Also
it is clear that the cloud has a much sharper edge facing VLA1, suggesting
that the stellar jet has eroded away one side of the cloud.

\epsfverbosetrue
\epsfxsize=8.0 cm
\begin{figure}
\center{
\leavevmode
\epsfbox{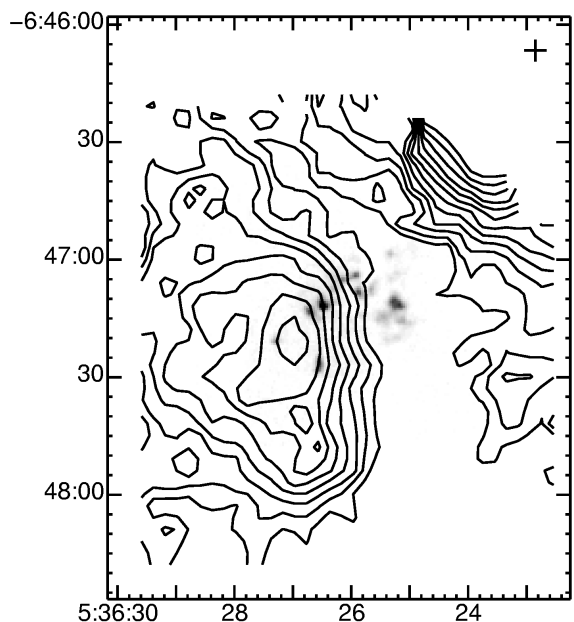}
\caption { Contours of sub-mm continuum emission in the vicinity of HH2, 
superimposed on the near-infrared H$_2$ image from Davis et al. (1994).
Contour interval is 22~mJy/beam, with lowest contour 44 and peak of
200~mJy/beam.
The beam size in the sub-mm data is 14$\arcsec$ fwhm.
VLA1, the driving source for the flow, is shown by the cross in the upper
right; the outer edge of the dust disc associated with this young
star can be seen towards the upper right of the contour map.
The map coordinates are J2000.0.
}
}
\end{figure}

The total integrated 850$\micron$ flux from the extended cloud associated with
HH2 is estimated at 2.1$\pm$0.4~Jy (this is measured over a region
of $80\asec \times 60$ arcsec at $PA=30^{\circ}$ centred on the peak).
The peak flux is 0.20$\pm$0.04~Jy, consistent
with an earlier single point photometric observation at 800$\micron$
(Dent, 1997). By comparison, the exciting star VLA~1 has a peak flux
of 3.3~Jy (Chini et al., 2001). Both HH2 and VLA~1 show a compact
central source superimposed on an extended low-level plateau;
in both cases the latter may be heated by the ambient interstellar
radiation field rather than the central
source (see Chini et al., 2001).

Pravdo \& Chester (1987) detected emission in the region of HH2
at 12 and 25$\micron$ using IRAS, which they tentatively
ascribed to the HH object. However, their maps indicated
the mid-infrared peak may be $\sim 1 \amin$ west of HH2;
this is supported
by more recent ISO data (Cernicharo et al., 1999), suggesting the
mid-infrared continuum is concident with the H$_{\alpha}$ rim
noted by Reipurth et al. (1993) to lie 50$\asec$ west of HH2.
However, far-infrared ISO results did
show continuum emission from the region of HH2 itself (Molinari \&
Noriega-Crespo, 2002); the estimated contamination
from the bright source VLA~1 was $\leq $10\%.
Figure~3 shows the integrated continuum fluxes.
A grey body fit to the long-wavelength data 
gives a temperature of 22$\pm$2~K and
dust opacity index, $\beta$=1.5$\pm$0.2; this would imply a total
cloud luminosity of
13$\pm 5 \Lsun$. The mid-IR points are consistent with
T=220K, but these are ignored in this and subsequent analysis for reasons
given above. Furthermore the temperature of this component is 
sufficiently high that emission is unlikely to significantly affect 
the far-infrared fluxes and hence the fit (see Fig.~3).
However, further high-resolution imaging in both the
far and mid-infrared is required to confirm this result.

\epsfverbosetrue
\epsfxsize=8.0 cm
\begin{figure}
\center{
\leavevmode
\epsfbox{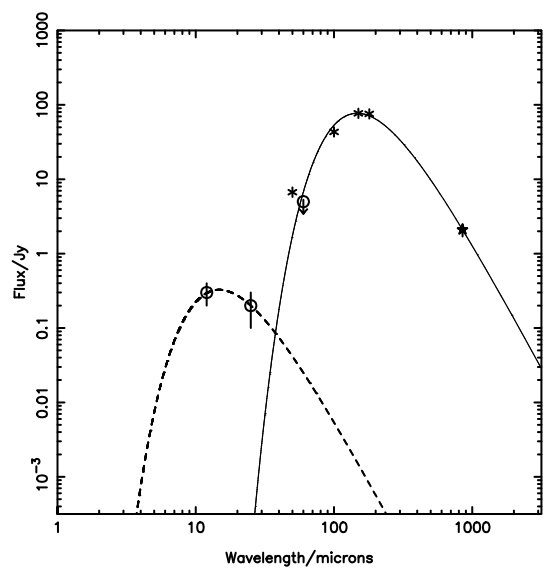}
\caption {Spectral energy distribution of the cloud near
HH2. The 850$\micron$ point represents
the total emission from the cloud associated with HH2 (see text).
Data points indicated by stars
are based on results from Molinari \& Noriego-Crespo (2002),
and open circles are
IRAS data points taken from Pravdo \& Chester (1987). The solid
line shows a fit to the long-wavelength data only, with T=22K, $\beta$=1.5.
The 12 and 25$\mu m$ points are fitted assuming the same
value of $\beta$, and with T=220K; however, this mid-infrared emission may
arise from a nearby source not directly associated with HH2 - see text for
details.}
}
\end{figure}

The NH$_3$ observations of Torrelles et al.
(1992) indicated kinetic temperatures in the region of the HH2 clump
of $\leq$20K. Girart et al. (2002) derive CO excitation temperatures of
$\sim$13K. The peak brightness temperature in $^{13}$CO and $^{12}$CO 
is $\sim$10K (see below), which is a lower
limit to the kinetic temperature assuming the line to be
optically thick and in LTE.

If we adopt a temperature of 22K, the {\em total} mass of the HH2 cloud
is 3.8$\pm 0.4 M_{\odot}$, assuming a gas:dust ratio of 100, and an
850$\micron$ dust mass opacity of $10^{-3} m^2 kg^{-1}$ (eg Henning et
al., 1995). This gives a mean gas column density of
$1.8 \pm 0.2 \times 10^{22} cm^{-2}$ and mean space density of
$5 \pm 0.5 \times 10^4 cm^{-3}$ (assuming a symmetrical cloud).
Note that the errors on these derived parameters may be larger
- perhaps up to a factor of 2 - because of the uncertainty in the
mm-wavelength dust emissivity (eg Henning et al.).

The contours in Figure~2 show a clear emission peak near HH2;
assuming the temperature and dust opacity
here is the same as that of the whole cloud, then
the luminosity of this component is estimated as $\sim 1.2 \pm 0.3 \Lsun$.
Its mass is $0.3 \pm 0.08 \Msun$, peak column density
$2.3 \pm 0.6 \times 10^{22} cm^{-2}$
and peak space density is $2 \pm 0.5 \times 10^5 cm^{-3}$.
These densities are very similar to those found from recent observations
of several molecular species towards HH2
(Girart et al., 2002), although the derived temperature is somewhat higher.
This general agreement confirms that the dust
and molecular gas lie in the same warm clump, and also suggests that
the sub-mm dust emission characteristics
are not significantly affected by the nearby shock.

\subsection{Single-dish CO}

The outflow axis from VLA1 lies close to the plane of the sky, resulting in
relatively faint CO line wings. Nevertheless a collimated
molecular outflow has been observed close to VLA1 (Correia et al., 1997;
Choi \& Zhou, 1997), and fainter red-shifted CO extends as far as HH2
(Moro-Mart\'in et al., 1999).  In an attempt to understand this termination
region, we observed HH2 itself using the J=3-2 transition of CO and its
isotopomers; the higher energy level of this transition compared
with earlier observations means it is more sensitive to the warmer gas.
Figure~4 compares spectra at the location of the continuum dust peak. 
The C$^{18}$O line can be fit by a single
Gaussian component at $7.25 \pm 0.2 km s^{-1}$, close to that of
the HCO$^+$ (7.0$km s^{-1}$; Davis et al., 1990),
CI (Dent, 1997), and other species such as SO (Girart et al., 2002). 
This velocity is significantly different from the gas around
VLA1, which has a radial velocity of 10.5 $km s^{-1}$
(Choi \& Zhou 1997).

The ratio of peak brightness temperature of $^{13}$CO to C$^{18}$O
is $\sim$5 which, assuming a similar emitting area and LTE
conditions, indicates that the C$^{18}$O line is optically thin.
The integrated C$^{18}$O intensity is $2.6 K km s^{-1}$; with an excitation 
temperature of 22K (see above) and a
C$^{18}$O abundance of $10^{-7}$, this gives a total gas column density
of $1.4 \times 10^{22} cm^{-2}$. This is consistent with the value
estimated above from the sub-mm dust emission.

\epsfverbosetrue
\epsfxsize=8.0 cm
\begin{figure}
\center{
\leavevmode
\epsfbox{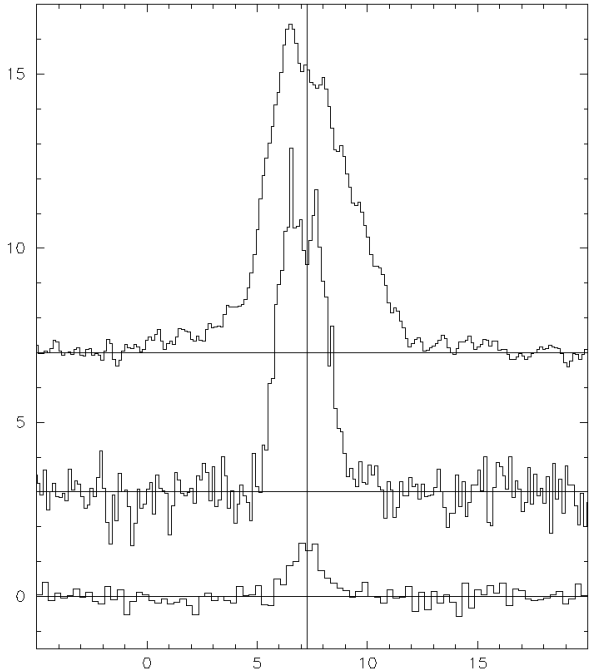}
\caption { Spectra of CO J=3-2 towards the continuum peak near HH2.
From top to bottom, these are $^{12}$CO,
$^{13}$CO and C$^{18}$O. Intensity scale is
T$_{mb}$, and velocity (in $km s^{-1}$) is with respect to {\it lsr}.
The vertical line represents the systemic radial velocity of the
molecular clump associated with HH2.}
}
\end{figure}

The $^{12}$CO line in Figure~4 shows distinct high-velocity wings
at relative velocities $|v_{rel}| \geq 3 km s^{-1}$.
The spatial distribution of this high-velocity
gas in the vicinity of HH2 is illustrated
in Figure~5; both the red and blue-shifted emission
peak within 10$\asec$ of the brightest H$_2$ knot HH2E.
A second component of blue-shifted gas lies south of HH2L. 
It was shown by Moro-Mart\'in et al. that red-shifted CO around HH1-2
traces a large
collimated flow centred on VLA1; high-resolution observations of
blue-shifted HCO$^+$ (to be described below) show that it too lies
in a collimated jet pointing back towards VLA1.
It is thought unlikely that the dust clump in Fig.~2 harbours a
young outflow source, as the low extinction through the cloud
would render it detectable in the infrared.
So the abrupt termination of the outflow near HH2E (Fig.~5),
suggests this is the dominant interaction region of the jet and ambient clump.
However the spatial resolution of these single-dish data is insufficient
to compare accurately with the infrared image.

\epsfverbosetrue
\epsfxsize=6.0 cm
\begin{figure}
\center{
\leavevmode
\epsfbox{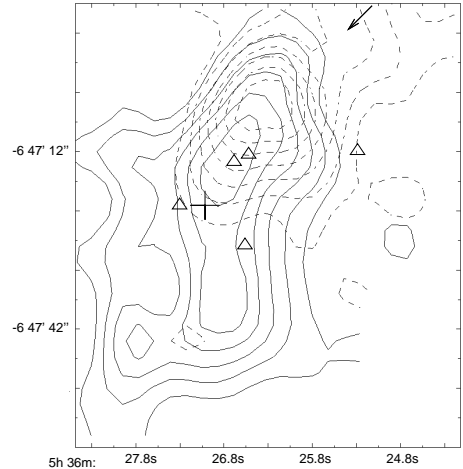}
\caption { Distribution of high-velocity red- and
blue-shifted $^{12}$CO J=3-2 emission in the vicinity of HH2. The
dashed contours delineate red-shifted gas integrated over the range
13 - 18 $km s^{-1}$ and the solid contours show gas in the range
0 - 5 $km s^{-1}$. Contour interval is
0.8 $K km s^{-1}$, with a base of 0.8 (temperature scale in antenna units).
The coordinates are J2000.0; beam size is $14\asec$ (fwhm).
The location of some of the major HH clumps are indicated by
triangles; from the South these are HH2L, Knot 8, HH2F, E and A.
The peak of the sub-mm continuum is shown by a cross and the presumed
direction of the outflow from VLA1 is shown by the arrow.}
}
\end{figure}

\subsection{Interferometric HCO$^+$}

The HCO$^+$ J=1-0 emission from the vicinity of HH2 is bright and complex, but
the spectra have two distinct velocity components. Firstly,
bright and relatively narrow ambient-velocity emission
with a profile similar to that of $^{13}$CO; this is illustrated in
Figure~6a which shows the line
towards the continuum peak. Secondly a high-velocity wing
component, with
a full width of 20 $km s^{-1}$ (to the noise level). This dominates
in the region of HH2E (see Figure~6b).

\epsfverbosetrue
\epsfxsize=8.0 cm
\begin{figure}
\center{
\leavevmode
\epsfbox{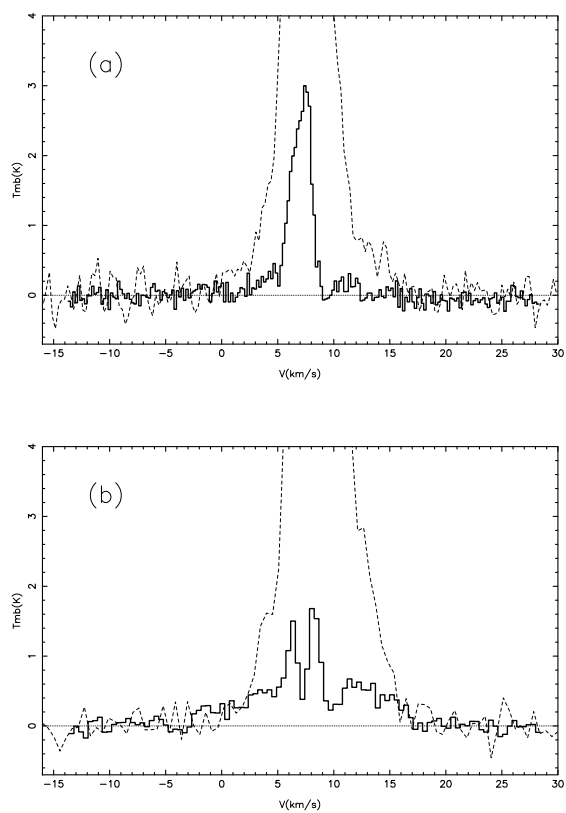}
\caption {Comparison of the J=1-0 HCO$^+$ line
(thick histograms) with J=3-2 $^{12}$CO (dashed lines) towards 
(a) the continuum peak and (b) HH2E.
High-velocity emission is seen in both
molecules, although the HCO$^+$ line is wider than CO. The HCO$^+$
data have been smoothed to the same resolution as the single-dish data
(14$\asec$); intensity scale is T$_{mb}$ and velocity frame is {\em lsr}.
}
}
\end{figure}

Both HCO$^+$ spectra show an absorption dip at
$v_{lsr} \sim 10 km s^{-1}$, similar to the velocity of the
molecular gas near VLA~1 and in much of the Orion region
(Davis et al., 1990; Choi \& Zhou, 1997). It is likely that a cool region
of this cloud along the line of sight is absorbing the emission from HH2.
However, a second dip in the HCO$^+$ spectrum
at $v_{lsr}\sim 7 km s^{-1}$ is likely to be cool foreground gas
associated with the HH2 clump itself,
as seen in the CO spectra in Fig.~4.

\epsfverbosetrue
\epsfxsize=8.0 cm
\begin{figure}
\center{
\leavevmode
\epsfbox{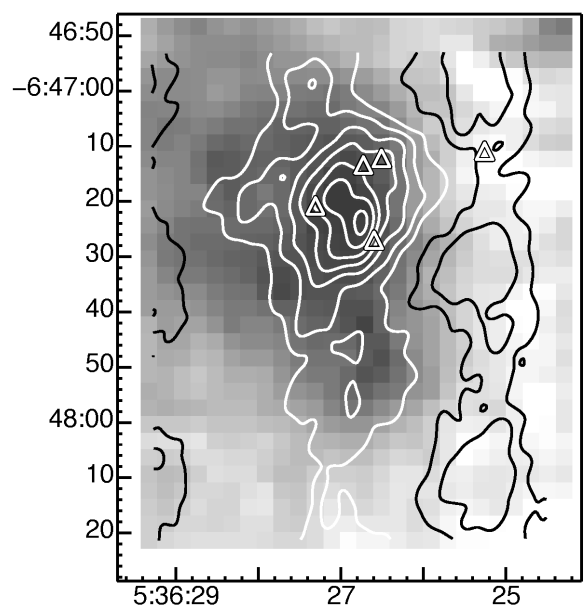}
\caption {Contours of ambient-velocity HCO$^+$ (5.0 to 8.5 $km s^{-1}$)
superimposed on a greyscale image of the
sub-mm continuum.  Contour levels are -8, -4, 4, 8,
12 ... $K km s^{-1}$; negative contours are shown in black. Note that
the negative
contours are due to the limited UV coverage of the observations; they
are not thought to significantly affect the morphology of the peak
emission.
Also shown by triangles are the locations of the HH objects HH2L, Knot
8, HH2F, HH2E, and HH2A (in order, from the South). Coordinates are J2000.
The synthesised beam size is $6.9\asec \times 4.6\asec$ at $PA = -20^{\circ}$.
}
}
\end{figure}

The map of ambient velocity HCO$^+$ (Figure~7) shows a main peak of size
$20 \asec$ (fwhm), consistent with the lower-resolution observations
of Davis et al. (1990). Fig.~7 compares the data with the sub-mm
continuum emission and the locations of some of the prominent HH
knots (c.f. Fig.~1). The gas and
dust distributions are similar but not identical; notably the HCO$^+$ shows
two compact peaks within
$5 \asec$ (0.01pc) of
HH2L and Knot 8, whereas the dust is seen in a broader North-South
ridge extending over $\sim 50 \asec$. It is possible
that some of the large-scale
HCO$^+$ emission ($\geq 52 \asec$) has been resolved out by the lack of
short-baseline data from the interferometer. Additional instrumental
effects due to incomplete UV coverage can be seen as negative contours
in Figure~7, although these are at a relatively low level compared with the
peak intensity. The map may also be affected by the self-absorption near
the line centre, although this is relatively narrow compared with the
main emission profile. Despite these potential difficulties,
it does appear that the ambient-velocity HCO$^+$ emission closely
traces the dust continuum; however, it appears relatively brighter
within 0.01pc of the HH shocks HH2L and Knot 8.
This enhancement will be discussed further in Section 4.1.

\subsection{High-velocity HCO$^+$}

Maps of
the integrated high-velocity HCO$^+$ emission are shown in Figure~8, with
the near-infrared H$_2$ image (Davis et al., 1994)
superimposed. The distribution of
high-velocity gas appears similar to that of 
CO shown above, after accounting for the differing resolution and relative
insensitivity of the interferometric observations
to the large scale structure ($\geq 52 \asec$).
High-velocity red and blue-shifted
HCO$^+$ is found within 5$\asec$ of the HH objects HH2K, E, F and also Knot 8.
Notably, these knots are relatively bright in shocked H$_2$ compared
with optical lines such as H$_{\alpha}$, whereas HH2A and HH2H, which dominate
optical images, are relatively faint in H$_2$ (eg Noriego-Crespo \&
Garnavich, 1994). Multi-level line analysis shows
that they also have lower excitation temperatures
(Eisl\"offel et al., 2000). There is weak evidence of
high-velocity red-shifted gas near the western HH clump HH2A.
 
The images of high-velocity HCO$^+$ (particularly the blue-shifted gas)
show that the northern section of
HH2, rather than being a jumble of shocks,
actually forms a rather coherent well-collimated jet. This includes
HH2K, E, and F, terminating in Knot 8 (and thus confirming
that the latter object, which is only detected in H$_2$, is part
of the flow).
The Position Angle of this jet measured from VLA1
is 139$^{\circ}$ - significantly different from the $PA$ of
the optical peak HH2A measured from VLA1 (150$^{\circ}$). By comparison, the
apex of the large-scale
outer bow shock HH402 lies at $PA \approx 138^{\circ}$
(Ogura et al., 1995). Assuming a constant flow direction, these
results suggest that the dominant outflow position angle is 138$^{\circ}$.
This is further confirmed by the fact that the sub-mm
continuum peaks at $PA \sim 140^{\circ}$ from VLA1; if we
assume it is heated by the jet, this traces where most of the outflow energy
is being deposited.

Giarart et al. (1999) suggested that HCO$^+$ may be used to identify
shocked gas in sections of outflows that are no
longer visible in optical or infrared lines.  Note that the cooling times
associated with the optical/IR lines are very short - of the order of a
few years - so these emission lines can rapidly fade.  The HCO$^+$ lines, on
the other hand, emit from molecular material that is already relatively
cool, where
steady depletion via dissociative recombination with electrons may take 
$ \geq 10^2$ years (Nejad \& Wagenblast 1999), although this is dependent
on the electron abundance.  Indeed, the apparent
HCO$^+$ abundance enhancement evident in our data may be 
limited by the lifetime of the clump in HH2 which, if we assume
is destroyed by the passage of the shock front,  
will be of the order of a few thousand years.

\epsfverbosetrue
\epsfxsize=8.0 cm
\begin{figure}
\center{
\leavevmode
\epsfbox{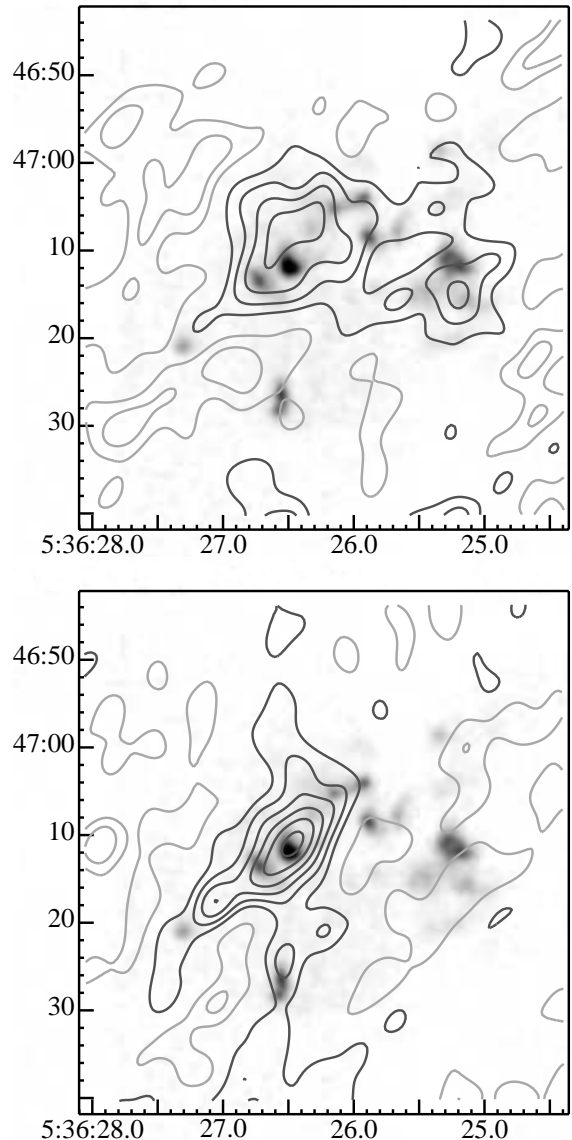}
\caption {Contours of high-velocity red (upper panel) and blue-shifted HCO$^+$
superimposed on the H$_2$ image. Red-shifted gas is integrated over
+11.0 to +20.0 $km s^{-1}$ and blue-shifted gas from -5 to +5.0 $km s^{-1}$.
Contours are -0.4, -0.2, 0.2, 0.4, 0.6... $K km s^{-1}$.
Map coordinates are J2000.
The synthesised beam size is $6.9\asec \times 4.6\asec$ at $PA = -20^{\circ}$.
}
}
\end{figure}

\section{Discussion}

\subsection{The ambient clump}

Although the sub-mm results show an extended cloud
downwind of the bright optical HH objects, the central
sub-mm peak coincides with a number of
low-excitation HH knots. The cloud has no
evidence of an internal young star, so how is the dust heated? Either it
could be through the interstellar radiation field in the young cluster,
by UV radiation from optically-bright knots such as HH2A, or by
shock heating.
The fact that the submm continuum peaks so close to
HH2 suggests the heating - at least of this central peak - is local,
rather than from the external interstellar radiation field.
One possible local energy source is UV from the HH shocks,
as used to explain the HCO$^+$ overabundance in the clump (eg
Wolfire \& K\"onigl, 1993; Taylor \& Williams, 1996;
Viti \& Williams, 1999). B\"ohm-Vitense et al. (1982)
found the dominant UV source in HH2 lies within $\sim 6 \asec$ of the optically
bright high-excitation knots HH2H and A. Thus the main UV source
is $\sim 25\asec$ Northwest of the sub-mm peak, and
the fraction of UV radiation intercepted by the dust clump would
be $\sim 5$\%. The total de-reddened UV luminosity of HH2A,
estimated from Fig.~2 of B\"ohm-Vitense et al. is $\sim 0.3 \Lsun$, which
would result in an intercepted luminosity at the sub-mm peak
of $\sim 0.015 \Lsun$,
considerably less than that observed ($\sim 1.2 \Lsun$).
Clearly the UV flux from HH2H/A is inadequate to heat the dust peak.

Alternatively the clump may be warmed by local shocks from the outflow
jet. Moro-Mart\'in et al. give the total outflowing mass from VLA1
as 0.2$\Msun$; with a mean flow velocity of
$30 km s^{-1}$ (deprojected assuming an inclination of 10$^{\circ}$)
and age of $10^4 yrs$, this results in a total flow mechanical
luminosity of $\sim 1.0 \Lsun$. This is
adequate to explain the local heating of the sub-mm peak, although
not of the whole cloud shown in Fig.~2.

Both the sub-mm continuum and the ambient-velocity
HCO$^+$ reach a maximum within $5\asec$ (0.01~pc) of HH2L and Knot 8.
To estimate the HCO$^+$ enhancement in this region, we can compare
the molecular abundance
with a potentially more ``benign'' region of the cloud $25\asec$
to the South. Data are smoothed to the same spatial resolution
($14 \asec$ fwhm), and we make an initial assumption of
an excitation temperature
of 22K (see above), LTE conditions and optically thin emission.
Towards the dust peak and southern region, the HCO$^+$ column
densities ($N_L(HCO^+)$) are $3.6 \times 10^{13}$ and $1.1 \times 10^{13}
cm^{-2}$ respectively. This compares with estimates derived from
the dust of $N_L(H_2) = 2.3 \times 10^{22}$ and
$1.8 \times 10^{22}$, giving
abundances, $\chi_{HCO^+}$ of $1.6 \times 10^{-9}$
and $6.5 \times 10^{-10}$. Girart et al. (2002)
used multi-transition modelling to show that the HCO$^+$
excitation temperature may be somewhat less than 22K, possibly because
the mean cloud density is less than the critical density of $10^6 cm^{-3}$.
Furthermore they suggest the line may be optically thick towards the
bright peak, which is supported by the self-absorbed profile in Fig.~6a.
But if we assume similar conditions apply in the two regions,
it indicates $\chi_{HCO^+}$ in the
ambient gas is a factor of $\geq$2.5 higher within
$\sim$ 0.01~pc of HH2L, compared with that 0.05~pc to the South.
Furthermore, the abundance in the Southern region is similar to that in
many other quiescent molecular clouds (eg Nejad \& Wagenblast, 1999).

\subsection{High-velocity molecular gas}

High-velocity HCO$^+$ and CO emission is
closely associated with a line of H$_2$ knots terminating at Knot 8 (Fig. 8).
This suggests a single coherent jet 
impacting the dust clump, rather than a broad wind with multiple shocks
as might be suggested by the complex optical images.
A similar spatial association of high-velocity HCO$^+$ with H$_2$ peaks
was found in NGC2071 (Girart et al., 1999) and DR21 (Garden \&
Carlstrom, 1992), and it was proposed that the HCO$^+$ enhancement is due
to ion-chemistry reactions in the low-velocity C-shocks.

After accounting for the differing resolution of the data,
there appears to be no significant difference in the spatial
distribution of high-velocity CO and HCO$^+$.
However, there is a clear difference
in the line brightness ratio as a function of velocity, illustrated
in Figure~9. We find this ratio changes from
$\sim$0.1 at the lowest relative velocities likely to be uncontaminated by
ambient emission ($|v_{rel}| \sim 3 km s^{-1}$), up to $\sim$2-3 at the
highest velocities ($|v_{rel}| \sim 12 km s^{-1}$). Here we use
$v_{rel} = v_{lsr} - v_{sys}$, where $v_{sys}$, the systemic clump velocity,
is taken to be $7.0 km s^{-1}$.
Note that the inclination of the HH1-2 flow is only $\sim 10^{\circ}$ to
the plane of the sky (eg Moro-Mart\'in et al., 1999), so
the absolute velocities may be considerably higher. However,
the near coincidence of red and blue-shifted gas (Fig.~8) suggests
that random turbulent or tranverse velocities in the entrained gas
dominate the measured velocity.
As the line wings of both species are thought to be optically thin,
Fig.~9 suggests a factor of 30 variation in the relative abundance.
However, there are other possible explanations: Girart et al. (1999)
observed a similar factor of 10 variation in the HCO$^+$ to $^{12}$CO J=2-1
ratio across the NGC2071 outflow spectra. They discussed several
possible explanations, for example
changing excitation conditions, but concluded that an increase in
HCO$^+$ abundance was the most likely cause.

If we assume a constant excitation temperature of 30K and
the gas
is in LTE, then the results indicate $\chi_{HCO^+}$ varies from
$2 \times 10^{-8}$ up to $6.8 \times 10^{-7}$ at the extreme velocities.
This assumes the canonical value for $\chi_{CO} = 5 \times 10^{-5}$ applies
throughout the gas.
It represents a factor of up to $10^3$ increase in the abundance compared with
the ambient gas in the HH2 dust clump (see previous section).
If the excitation temperature is 100K, this
enhancement may be as high as $10^4$.
The optical extinction through dust mixed with the high-velocity gas
is likely to be negligible, so external UV irradiation cannot
easily explain the velocity-dependence of the HCO$^+$ enhancement.
However, Taylor \& Raga (1995) predicted HCO$^+$ abundances of up to
$10^{-6}$ in turbulent mixing layers associated with
relatively low-velocity ($40 km s^{-1}$) shocks. This could explain both
these enhancements, as well as the close association
of the high-velocity regions with the low-excitation, H$_2$-dominant knots.

\epsfverbosetrue
\epsfxsize=8.0 cm
\begin{figure}
\center{
\leavevmode
\epsfbox{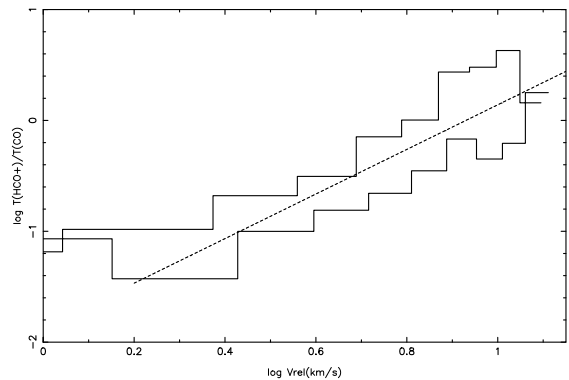}
\caption {Ratio of the J=1-0 HCO$^+$ to J=3-2 $^{12}$CO line intensity
in the region of HH2E ($5^h 36^m 26.52^s, -6^{\circ} 47' 12''$),
as a function of relative velocity,
for the blue (upper histogram) and red-shifted (lower histogram) gas.
Also shown is a fit to the average
of the blue and red-shifted components; the slope of this
line is 2.0.
The spectra were obtained by smoothing both datasets
to the same angular resolution (15$\asec$) and spectral channel width
(1.3$km s^{-1}$). Only the velocity regions with significant
detections of both lines are shown.
}
}
\end{figure}

The brightness ratio in Fig.~9 shows a monotonic
increase with velocity, consistent with $\chi_{HCO^+} \propto v_{rel}^2$.
HCO$^+$ line wings in several outflows are found to have
a relatively constant or even increasing brightness
temperature out to the highest velocities (eg HH7-11 - Lizano et al., 1988;
L1551 - Rudolph, 1992; NGC2071 - Girart et al., 1999; OH231.8 -
S\'anchez-Contreras et al, 2000).
By contrast, the intensity of CO line wings are generally
found to decrease with $v_{rel}^{-1.8}$ (eg Richer et al., 2000), suggesting
that the $v^2$
dependence of $\chi_{HCO^+}$ may be a common feature of shocked and entrained
gas in such outflows.
As depletion of HCO$^+$ occurs on a timescale significantly longer
than the age of the flow, then the velocity-dependent abundance
suggests either that: (1) a range of shock
velocities exists, with the HCO$^+$ enhancement depending on the
shock kinetic energy, or (2) HCO$^+$ is formed
at the higher shock velocities, and this gas gradually mixes
with ambient un-enhanced material. The latter might take place in a
steady-state turbulent boundary layer (eg Taylor \& Raga, 1995). We can
use the results to estimate $\epsilon (v)$, the mixing fraction in the
turbulent boundary, ie the ratio
of masses of initially shocked gas, $m_s(v)$, to ambient gas, $m_a(v)$, in this
turbulent boundary. The HCO$^+$ results above indicate that $m_a(v)$ is
approximately constant, whereas CO data show approximately
that $m_s \sim v^{-2}$.
This would imply $\epsilon (v) \sim (v/v_j)^2$, where $v_j$ is the maximum
HCO$^+$ velocity. Further modelling of such low-velocity turbulent
layers would be of interest.

\subsection{HCO$^+$ enhancement and shocks in the clump}

The relatively modest enhancement of the HCO$^+$ abundance
towards the ambient velocity HH2 clump has been
ascribed to a radiative precursor from the optically-bright
HH knots HH2A/H (eg Wolfire \& K\"onigl, 1993; Raga \& Williams,
2000). However, the current results (Fig.~7) do not show the limb-brightened
morphology in HCO$^+$ predicted by Raga \& Williams, even
though the flow axis lies close to the plane of the sky.
Instead the results suggest a closer association with the
low-excitation shocks, found by H$_2$ and high-velocity HCO$^+$
within $\sim$ 0.01pc of the clump. Luminosity arguments suggest
the dust clump is heated by the flow itself. Furthermore, the dust
morphology indicates that one side of the clump has been
truncated by the flow impact (Fig.~2).
Clearly, then, at least some parts of the clump have been dynamically affected
by the outflow. 
It is perhaps worth noting that low-velocity HCO$^+$ clumps in the
outflow from NGC2264G are also closely associated with low-energy
H$_2$ shocks and accelerated gas (Girart et al., 2000).
So could the enhancement of HCO$^+$ at velocities close
to ambient be due to low-velocity shocks within the dust clump,
in a similar mechanism as evoked to explain the high-velocity enhancement?

The HCO$^+$ linewidth in the HH2 clump
is similar to that of the bright cores associated
with VLA1 and VLA3 (eg Choi \& Zhou, 1997), even though the mass is
10-100 times lower. We can estimate the
ratio $2 K / P$ in the HH2 and VLA1 clumps, where $K$ and $P$ are
the kinetic energy (assumed due to turbulence) and the gravitational
potential energy. Using the
masses and core sizes from section 3.1 above, and from Choi \& Zhou, we find
ratios of $65$ and 0.6 for HH2 and VLA1, showing that, unlike VLA1, 
the clump near HH2
is clearly not gravitationally bound. One possible mechanism for the
line broadening is disruption by the HH2 jet. In which case
the bright ambient-velocity HCO$^+$ could also be caused by
enhanced abundance in the lowest velocities of the turbulent mixing layer.
Extrapolating the $v^2$ enhancement seen in the high-velocity gas
down to $v_{rel} \sim 1 km s^{-1}$ we would predict abundances a factor
of $\sim 10$ above normal, similar to that observed.

\section{Conclusions}

A sub-mm continuum clump is found at the end of the molecular outflow and
0.05pc downwind of the optically-bright knots in HH2.
Warming of the dust appears to be caused by the impact of the jet, and
the derived luminosity of the central peak ($\sim 1.2 \Lsun$)
is similar to that of the outflow mechanical luminosity.

The emission from HCO$^+$ can be divided into an ambient
velocity component, approximately correlated with the dust emission,
and a high-velocity component, closely associated with the H$_2$ knots.
The high-velocity HCO$^+$ appears to form a single coherent
flow linking the lower-excitation HH knots; this is North of the
optically-bright regions, and we suggest it represents the main
collimated jet from VLA1. In the highest velocity line wings, HCO$^+$ is
enhanced by a factor of up to $\sim 10^3$,
compared with a factor of a few in the ambient-velocity gas. The enhancement
is found to increase as $v_{rel}^2$.

Enhancement of ambient-velocity HCO$^+$ abundance is most prominent within
0.01~pc of the low-excitation shocks at the tip of the outflow jet.
It is suggested that enhancement in this gas could be caused
by the same shock mechanism and turbulent mixing as used to explain
the high-velocity HCO$^+$. Thus a UV precursor may therefore not be necessary.

\section*{Acknowledgments}

The James Clerk Maxwell Telescope is operated by the Joint Astronomy
Centre on behalf of the United Kingdom Particle Physics and Astronomy
Research Council, the Netherlands Organisation for Scientific Research,
and the National Research Council of Canada. The
Nobeyama Radio Observatory is a branch of the National Astronomical
Observatory,
operated by the Ministry of Education, Culture, Sports, Science and Technology,
Japan. The authors thank S. Sakamoto for the support of our NMA
observations, and the referee for helpful comments.

\section{References}

\parindent 0mm

B\"ohm-Vitense, E., B\"ohm, K.H., Cardelli, J.A., Nemec, J.M., 1982, \apj,
262, 224

Cernicharo, J., Cesarsky, D., Noriego-Crespo, A., Lefloch, B.,
Moro-Mart\'in, A., 1999, in ``H$_2$ in Space'', ed., F. Combes, \& G. Pineau
des For\^ets, (Cambridge Univ. Press), 23

Chini, R., Ward-Thompson, D., Kirk, J.M., Nielbock, M., Reipurth, B.,
Sievers, A., 2001, \aa, 369, 155

Choi, M. \& Zhou, S., 1997, \apj, 477, 754

Correia, J.C., Griffin, M., Saraceno, P., 1997, \aa, 322, L25

Davis, C.J. \& Eisl\"offel, J., Ray, T.P., 1994, \apj, 426, L93

Davis, C.J., Dent, W.R.F., Bell Burnell, S.J., 1990, \mnras, 224, 173

Dent, W.R.F., 1997, in Malbet F., Castets A., eds., Poster Proc. IAU Symp.
182, Herbig Haro Objects and the Birth of Low Mass Stars, p. 88

Eisl\"offel, J., Smith, M.., Davis, C.J., 2000, \aa, 359, 1147

Garden R.P., Carlstrom, J.E., 1992, \apj, 392, 602

Girart, J.M., Ho, P.T.P., Rudolph, A.L., Estalella, R., Wilner, D.J.,
Chernin, L.M., 1999, \apj, 522, 921

Girart, J.M., Estalella, R., Ho, P.T.P., Rudolph, A.L., 2000, \apj, 539, 763

Girart, J.M., Viti, S., Williams, D.A., Estalella, R., Ho, P.T.P, 2002,
\aa, 388, 1004

Henning, Th., Michel, B., Stognienko, R., 1995, P \& SS, 43, 1333

Herbig G., Jones, B.F., 1981 \aj, 86, 1232

Hester, J.J., Stapelfeldt, K.R., Scowen, P.A., 1998, \aj, 116, 372

Lizano, S., Heiles, C., Rodr\'iguez, L.F., Koo, C.-C., Shu, F.H.,
Hasegawa, T., Hayashi, S., Mirabel, I.F., 1988, \apj, 328, 763

Molinari, S. \& Noriego-Crespo, A., 2002, \aj, 123, 2010

Moro-Mart\'in, A., Cernicharo, J., Noriego-Crespo, A., Mart\'in-Pintado, J.,
1999, \apj, 520, L111

Noriego-Crespo, A., Garnavich, P.M., 1994, \aj, 108, 1432

Nejad, L.A.M., Wagenblast, R., 1999, \aa, 350, 204

Ogura, K., 1995, \apj, 450, L23

Pravdo, S.H., Rodrigu\'ez, L.F., Curiel, S., Canto, J., Torrelles, J.M.,
Becker, R.H., Sellgren, K., 1985, \apj, 293, L35

Pravdo, S.H., Chester, T.J., 1987, \apj, 314, 307

Raga, A.C., Williams, D.A., 2000, \aa, 358, 701

Reipurth, B., Heathcote, S., Roth, M., Noriega-Crespo, A., Raga, A.C.,
1993, \apj, 408, L49

Richer, J.S., Shepherd, D.S., Cabrit, S., Bachiller, R., Churchwell, E.,
2000, Protostars and Planets IV, eds. Mannings, V., Boss, A.P., Russell, S. S.,
867, Univ. of Arizona Press

Rudolph, A., 1992, \apj, 397, L111

Rudolph, A., \& Welch, W.J., 1992, \apj, 395, 488

S\'anchez-Contreras, C., Bujarrabal, V., Neri, R., Alcolea, J., 2000, \aa, 
357, 651

Taylor, S.D., Raga, A.C., 1995, \aa, 296, 823

Taylor, S.D., Williams, D.A., 1996, \mnras, 282, 1343

Torrelles, J.M., Rodr\'iguez, L.F., Cant\'o, J., Anglada, G., G\'omez, J.F.,
Curiel, S., Ho, P.T.P., 1992, \apj, 396, L95

Viti, S., Williams, D.A., 1999, \mnras, 310, 517

Wolfire M.G., K\"onigl, A., 1993, \apj, 415, 204

\end{document}